\documentclass[prb,twocolumn,showpacs,preprintnumbers,amsmath,amssymb]{revtex4}
\usepackage{graphicx}
\usepackage{times}
\date{\today}
\begin{document}
 \title{Finite-density corrections to the Unitary Fermi gas: A lattice perspective from Dynamical Mean-Field Theory}
\author{Antonio Privitera$^{1,2}$}
\author{Massimo Capone$^{1}$}
\author{Claudio Castellani$^{1}$}
\affiliation{$^{1}$CRS SMC, CNR-INFM and Dipartimento di Fisica, Universit\`a di Roma ``La Sapienza'', Piazzale Aldo Moro 2, I-00185 Roma, Italy\\
$^{2}$ Institut f\"ur Theoretische Physik, Johann Wolfgang Goethe-Universit\"at, 60438 Frankfurt am Main, Germany}
%\author{Massimo Capone}
%\affiliation{CRS SMC, CNR-INFM and Dipartimento di Fisica, Universit\`a di Roma ``La Sapienza'', Piazzale Aldo Moro 2, I-00185 Roma, Italy}
%\author{Claudio Castellani}
%\affiliation{CRS SMC, CNR-INFM and Dipartimento di Fisica, Universit\`a di Roma ``La Sapienza'', Piazzale Aldo Moro 2, I-00185 Roma, Italy}

\begin{abstract}
We investigate the approach to the universal regime of the dilute unitary Fermi  gas as the density is reduced to zero in a lattice model.
To this end we study the chemical potential, superfluid order parameter and internal energy  of the attractive Hubbard model in three different lattices with densities of states (DOS) which share the same low-energy behavior of fermions in three-dimensional  free space: a cubic lattice, a ``Bethe lattice" with  a semicircular DOS, and a ``lattice gas" with parabolic dispersion  and a sharp
energy cut-off that ensures the normalization of the  DOS. The model is solved using Dynamical Mean-Field Theory, that treats directly the thermodynamic limit and  arbitrarily low densities, eliminating finite-size effects.  
At densities of the order of one fermion per site the lattice and its specific form dominate the results. The evolution to the low-density limit is smooth and it does not allow to define an unambiguous low-density regime.
Such finite-density effects are significantly reduced using the lattice gas, and they are maximal for the three-dimensional cubic lattice. Even though dynamical mean-field theory is bound to reduce to the more standard static
mean field in the limit of zero density due to the local nature of the self-energy and of the vertex functions, it compares well with accurate Monte Carlo simulations down to the lowest  densities accessible to the latter.
\end{abstract}
%opening

\maketitle

\section{Introduction}
The ever increasing ability to manipulate and control ultracold Fermi gases allows to experimentally realize in an accurate and controlled way physical conditions which can only be approximately realized in solid state,
 like the crossover from weak-coupling Bardeen-Cooper-Schrieffer (BCS) superfluidity to strong-coupling Bose-Einstein condensation (BEC) of preformed bosonic pairs, which occurs in two-component fermionic systems
 as a function of the coupling strength\cite{experimental_crossover}. 

Actual realizations of degenerate ultracold Fermi gases are dilute, i.e., the mean interparticle distance is much larger than the range $R_0$ of the interatomic potential. In terms of the Fermi momentum
$k_F=(3 \pi^2 n)^{\frac{1}{3}}$ the diluteness condition is equivalent to $k_F R_0 \ll 1$. For such dilute systems the effect of the interaction can be parameterized through the s-wave scattering
length $a_s$ only, regardless the details of the potential, provided its short-range character. This is particularly intriguing because  Fano-Feshbach resonances permit to control the scattering length
$a_s$ by simply tuning a magnetic field. One can thus move from negative values on the BCS side to positive values on the BEC side, passing through the resonance point, where $a_s$ diverges.
 This latter situation is usually referred to as unitary limit, and it displays an extra {\it universality} because the divergence of $a_s$ leaves us with a single length scale $\propto n^{-1/3}$ and a single energy scale, the Fermi energy\cite{HoPRL2004}. For example, the chemical potential and the superfluid energy gap will
 depend on the density only through the Fermi energy $E_F$, i.e. $\mu,\Delta \propto E_F$ though with non trivial coefficients. 
The evaluation of these coefficients defies simple analytical treatments, due to the strongly correlated nature of the problem, and to the lack of obvious expansion parameters.
 In this light numerical methods, like Quantum Monte Carlo (QMC) simulations\cite{carlson,astrak,bulgac1,bulgac2,burovski1,burovski2,burovski3,akkineni} of finite systems have
 an extremely important role, both to obtain direct estimates of the observables, and to guide the choice  of the relevant classes of diagrams in perturbative expansions.
 The use of QMC overcomes the limitations of perturbative methods, but QMC simulations still suffer from finite-size effects. Moreover they require a specific choice of the interaction potential, which may be relevant if the density is not small enough to assure the realization of the dilute limit. 

An alternative to a direct simulation of the dilute Fermi gas is to consider the low-density limit of three-dimensional lattice models with local interaction, like the attractive Hubbard model.
Obviously the lattice introduces a new length scale associated with the inter-site spacing. Therefore the universal behavior can only be recovered in the zero density limit,
where the infinite inter-particle distance makes the presence of the lattice irrelevant.
This approach has been used in the QMC simulations of Refs. \onlinecite{bulgac1,burovski1,burovski2,burovski3}, and it has
the advantage of an intrinsic regularization given by the lattice spacing, which from the diagrammatic point of view introduces an
ultraviolet cutoff. 
Nonetheless, even if QMC results for the attractive Hubbard model can
be regarded as essentially exact for finite systems, it should be kept
in mind that actual simulations are limited to a finite number of sites and the density can not be arbitrarily reduced.
 Hence they are plagued by both intrinsic finite-size effects and by finite-density effects,
 calling for careful extrapolations to the thermodynamic and zero-density limits.

In this paper we follow the lattice route to the dilute limit using  a theoretical approach which is explicitly built in the thermodynamic limit and it has no intrinsic limitations to treat arbitrarily small densities, the Dynamical Mean-Field Theory (DMFT)\cite{revdmft}. DMFT is a quantum, and more accurate, version of standard mean-field theories, in which the solution in the thermodynamic limit is obtained neglecting spatial correlations. This approach has been widely used in the context of solid-state physics\cite{revdmft}, and it proved accurate for most three-dimensional correlated solids, and more recently it has been proposed to access the properties of the dilute Fermi gas in Ref. \onlinecite{barnea}, where the system was studied in a three-dimensional cubic lattice. Since our focus is to understand the relevance of non-universal corrections to the universal regime, here we  extend the analysis of Ref. \onlinecite {barnea}  considering three different lattices (which will have different finite-density corrections), and much smaller densities.

Our analysis shows that the extrapolation to zero density requires extreme care. Even if the three lattices have the same low-energy density of states (which yields the same zero density limit), they provide significantly different results (i.e., they are not in a universal regime) down to densities of the order of 0.01 fermions per site. This is result is an important warning because present QMC calculations are limited to densities of this order of magnitude (to our knowledge the lowest density available for the present problem is $n = 0.05$ in Ref. \onlinecite{burovski3}).

The weakest dependence on density is found for a model, the so-called lattice gas, in which the density of states (DOS) coincides with that of fermions in free space up to a cut-off after which it vanishes. The results for the cubic lattice and for a lattice with a semicircular DOS are influenced to a larger extent by the large-density behavior. Only for extremely small densities
the three lattices provide the same result. 
Unfortunately we find that in the same small-density limit DMFT reduces to static mean-field (MF), since the method is unable to reproduce the divergence of $a_s$ beyond MF. 
However we will see that  the limitations of DMFT appear only at very small densities, while DMFT introduces important corrections to the static mean-field down to densities well below $n \simeq 0.05$.
This observation also explains why in Ref. \onlinecite{barnea} DMFT was found to extrapolate to a different value with respect to static MF. Indeed our analysis shows that in this reference the zero-density limit has been extrapolated from densities larger than those for which DMFT approaches static mean-field.

The paper is organized as follows: In Sec. II we introduce the attractive Hubbard model and the approach to the zero-density limit. Sec. III presents results of static mean-field; Sec. IV is devoted to the implementation of  Dynamical Mean-Field Theory for the attractive Hubbard model; Sec. V presents the DMFT results and Sec. VI contains the final remarks.

\section{Attractive Hubbard model approach to the unitary limit}

In this section we discuss how the three-dimensional attractive Hubbard model can be used to describe the properties of a two-component  dilute gas. The Hamiltonian of the model reads
 
\begin{equation}
\label{hubbard}
 \sum_{\vec{k},\sigma} (\epsilon_{\vec{k}}-\mu) n_{\vec{k},\sigma} + U \sum_i n_{i,\uparrow}n_{i,\downarrow},
\end{equation}

where $n_{k\sigma} = c^{\dagger}_{k\sigma}c_{k\sigma}$ ($n_{i\sigma} = c^{\dagger}_{i\sigma}c_{i\sigma}$) is the number operator for Fermions of spin $\sigma$ in momentum $k$ (site $i$).
$U<0$ is the strength of the local pairing interaction between fermions with different flavor, and $\epsilon_{\vec{k}}$ is the free dispersion in a chosen lattice with lattice spacing $l$.

The attractive Hubbard model has been studied in different frameworks as a paradigm for lattice superconductors and superfluids.\cite{review_attractive}
 Here we use this lattice model merely as a systematic way to approach the universal regime of an interacting gas. Therefore we focus on the small-density regime,
 where the effect of the lattice is bound to disappear because the average interparticle distance will ultimately become much larger than the lattice spacing.
 Nonetheless, at every finite density the details of the lattice are in principle relevant. 
 %The way in which these non-universal finite-density corrections disappear when the zero-density  limit is approached is the main subject of this work. 
 
 In our theoretical approach the lattice under consideration and its dispersion $\epsilon_{\vec{k}}$ enter the calculation only through the density of states. For a  system of non-interacting fermions with mass $m$ in three-dimensional continuum space the DOS per unit volume is (assuming $\hbar=1$ everywhere) $D_{free}(\epsilon)=\frac{2\sqrt{2}m^{\frac{3}{2}}}{4\pi^2}\sqrt{\epsilon}$. In order to obtain the correct zero density limit, we choose three lattices whose DOS's per lattice site $D_{lattice}$ share the same low-density behavior of $D_{free}$. This implies that $D_{lattice}(\epsilon) \simeq l^3D_{free}(\epsilon)$ for small energies, where $l$ is the lattice spacing that we will use as the unit length. In the following we drop the index ``lattice".

Namely we use:

\begin{itemize}
\item{A semicircular DOS
\begin{equation}
\label{scdos}
D_{SC}(\epsilon)=\frac{2}{\pi D} \sqrt{\frac{\epsilon}{D}(2-\frac{\epsilon}{D})}  
\end{equation}
 where $D$ is the half bandwidth, that we will use as energy unit in the rest of the manuscript. This DOS does not correspond to any obvious dispersion but, as we shall see, it has some practical advantages. 
 For small energies close to the bottom of the band we have $D_{SC}(\epsilon) \approx \frac{2 \sqrt{2}}{\pi D^{\frac{3}{2}}} \sqrt{\epsilon}$, which corresponds to $D_{free}$ if we  suitably choose the lattice effective mass $m=(4\pi)^{\frac{2}{3}}/D$.}

\item{The so-called lattice gas, whose DOS coincides with the free DOS up to a cutoff $\Lambda=1.40539 \ldots D$  above which it vanishes. This value of $\Lambda$ ensures the normalization of the DOS per lattice site. This model can be seen as an approximate version of a system in which only the momenta included in the Brillouin zone of the lattice are allowed, and for each of these momenta the dispersion coincides with the parabolic dispersion of free fermions electrons. Indeed implementing this procedure yields a similar DOS with a smoother energy cutoff. This model has been studied in Ref. \onlinecite{bulgac1,bulgac2,burovski3}, where a different value of the cutoff was used.}

\item{A three-dimensional lattice DOS for nearest-neighbor hopping on a cubic lattice with dispersion $\epsilon_{\vec{k}} = 
-2t(cosk_x+cosk_y+cosk_z)$ with $t=1/2m = D/2(4\pi)^{2/3}$. This choice has been used in Ref. \onlinecite{barnea} and \onlinecite{burovski1,burovski2}}

\end{itemize}

\begin{figure}[t]
\begin{center}
\includegraphics[width=7cm] {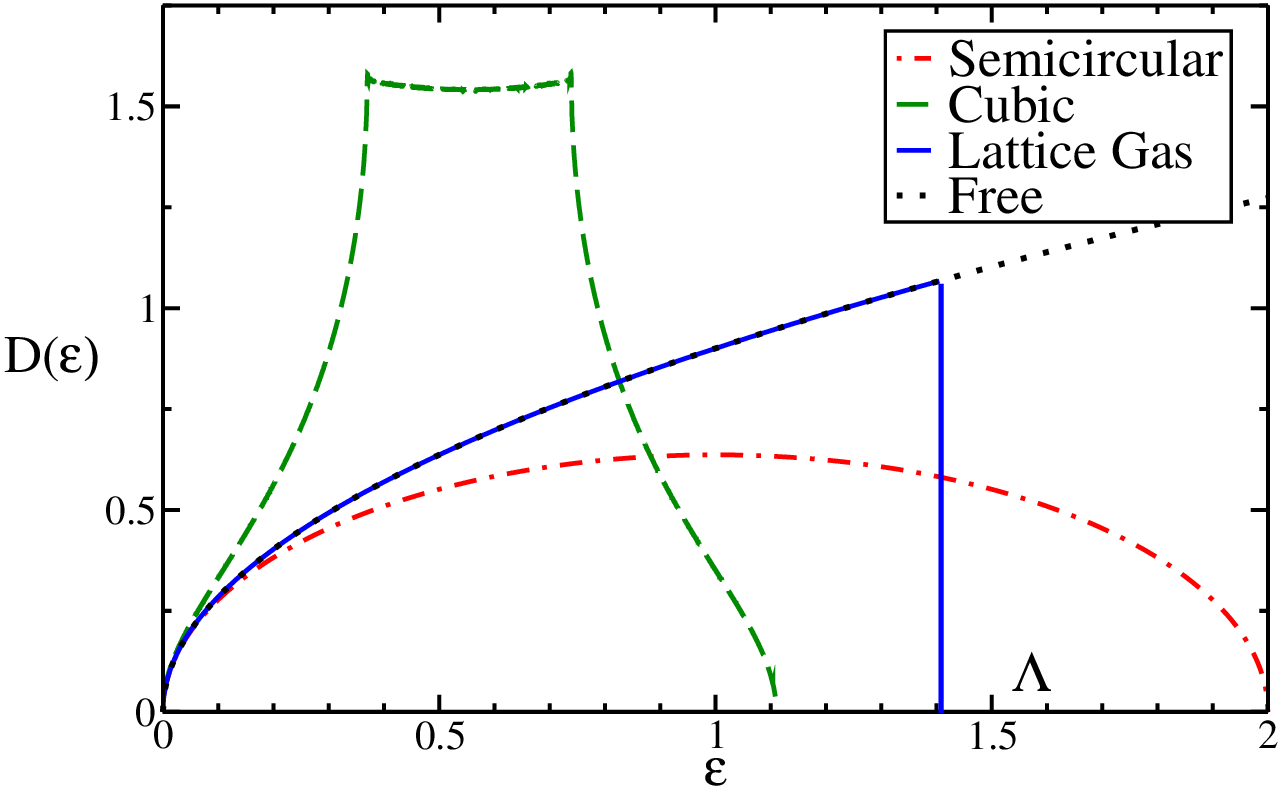}
\end{center}
\caption{\footnotesize{(Color Online) Comparison between the three different DOS's with parameters such that the low-energy behavior coincides with that of free electrons, as discussed in the text: semicircular DOS (dot-dashed red line), cubic three-dimensional lattice (dashed green line), lattice gas (solid blue line). For reference the DOS for   free fermions in three-dimensional space is shown as a dotted black line. Energies are expressed in units of the half-bandwidth $D$ of the semicircular DOS.}}
\label{figura1}
\end{figure}

In order to obtain a first insight on the properties of the different lattices, in  Fig. \ref{figura1} we compare the low-energy part of our DOS's 
(Obviously in this region the lattice gas coincides with the target DOS for fermions is three dimensions). It is interesting to observe that the semicircular DOS follows very closely the free 3-d DOS, and it is therefore likely to allow for a good convergence to the dilute limit $n\to 0$, while the cubic lattice rapidly departs from the reference.

\vspace{1cm}

As far as the interaction strength $U$ is concerned, in order to achieve the universal behavior in the dilute limit, $U$ has to be chosen as the unitary value,
for which the scattering length diverges.

For a Hubbard potential $a_s$ is related to the coupling constant $U$ through the relation\cite{burovski2}
\begin{equation}
 a_s(U)=\frac{m}{4 \pi \hbar^2} \frac{1}{U^{-1}+\Pi(\vec{0},0)}
\label{as_U}
\end{equation}
where
\begin{equation}
 \Pi(\vec{0},0)=\int_{-\infty}^{\infty} d \epsilon \ \frac{D(\epsilon)}{2\epsilon}=-U^{-1}_c
\end{equation}
For the semicircular DOS one finds $U_c=-D$, while for the lattice gas $U_c=-1.1624 \ldots D$ and for the cubic lattice $U_c = - 0.73224\ldots D$.

\section{A first insight from static mean-field}
\label{meanfield_sec}

In order to obtain a first feeling about the role of finite-density corrections to the dilute limit of lattice models and to understand the role of the precise choice of the lattice, we start from a simpler
approach with respect to DMFT, i.e. the static MF method. This approach shares the main advantages of DMFT with respect to exact numerical methods, being implemented in the thermodynamic limit without any restriction
in terms of attainable densities. Of course the method is less accurate than DMFT, which introduces exactly local quantum fluctuations, but it can be solved at very low computational cost at every density.
Finally, the zero-density limit of the method is the well-known Leggett's MF theory for the BCS-BEC crossover in dilute gases\cite{leggett_original}, which we can use as a benchmark to quantify finite-density
corrections and the dependence on the actual lattice.

The static MF approximation consists in decoupling the attractive interaction term both in the normal and in the anomalous s-wave Cooper channels,  determining the value of the pairing amplitude and of the chemical potential self-consistently. Then we introduce the scattering length using Eq. (\ref{as_U}). Finally we rescale the relevant energies by the non interacting Fermi energy $E_F$ of particles in $3d$ continuum space. The MF equations for the rescaled chemical potential $\tilde{\mu}=\mu/E_F$ and gap parameter $\tilde{\Delta}=\Delta/E_F$ are:

\begin{eqnarray}
 &&\frac{4}{3}=\int_{-\infty}^{\infty}\hspace{-0.5cm}d x\  \tilde{D}(x) \left[ 1 - \frac{(x-\tilde{\mu^\prime})}{\sqrt{(x-\tilde{\mu^{\prime}})^2 + \tilde{\Delta}^2}}\right] \label{mf1} \\
0=&&\hspace{-0.45cm}\frac{1}{k_Fa_s} \hspace{-0.1cm}=\frac{1}{\pi}\int_{-\infty}^{\infty}\hspace{-0.5cm} d x\  \tilde{D}(x)
\hspace{-0.1cm} \left [ \frac{1}{x} - \frac{1}{\sqrt{(x-\tilde{\mu^\prime})^2 + \tilde{\Delta}^2}} \right] \label{mf2}
\end{eqnarray}
where the left-hand side of Eq. (\ref{mf2}) vanishes because we take the unitary limit $a_s\to\infty$. 
$\tilde{D}(x)=\frac{D(E_F x)}{D_{free}(E_F)}$, $D(\epsilon)$ is the lattice DOS,
 $ \tilde{\mu^\prime} =  \tilde{\mu} - \frac{U_c n}{2 E_F}$ is the renormalized chemical potential including the Hartree contribution
 (which vanishes in the dilute limit, so it does not appear in Leggett's theory)  and $x=\epsilon/E_F$.
  
\begin{figure}[t]
\begin{center}
\includegraphics[width=7cm] {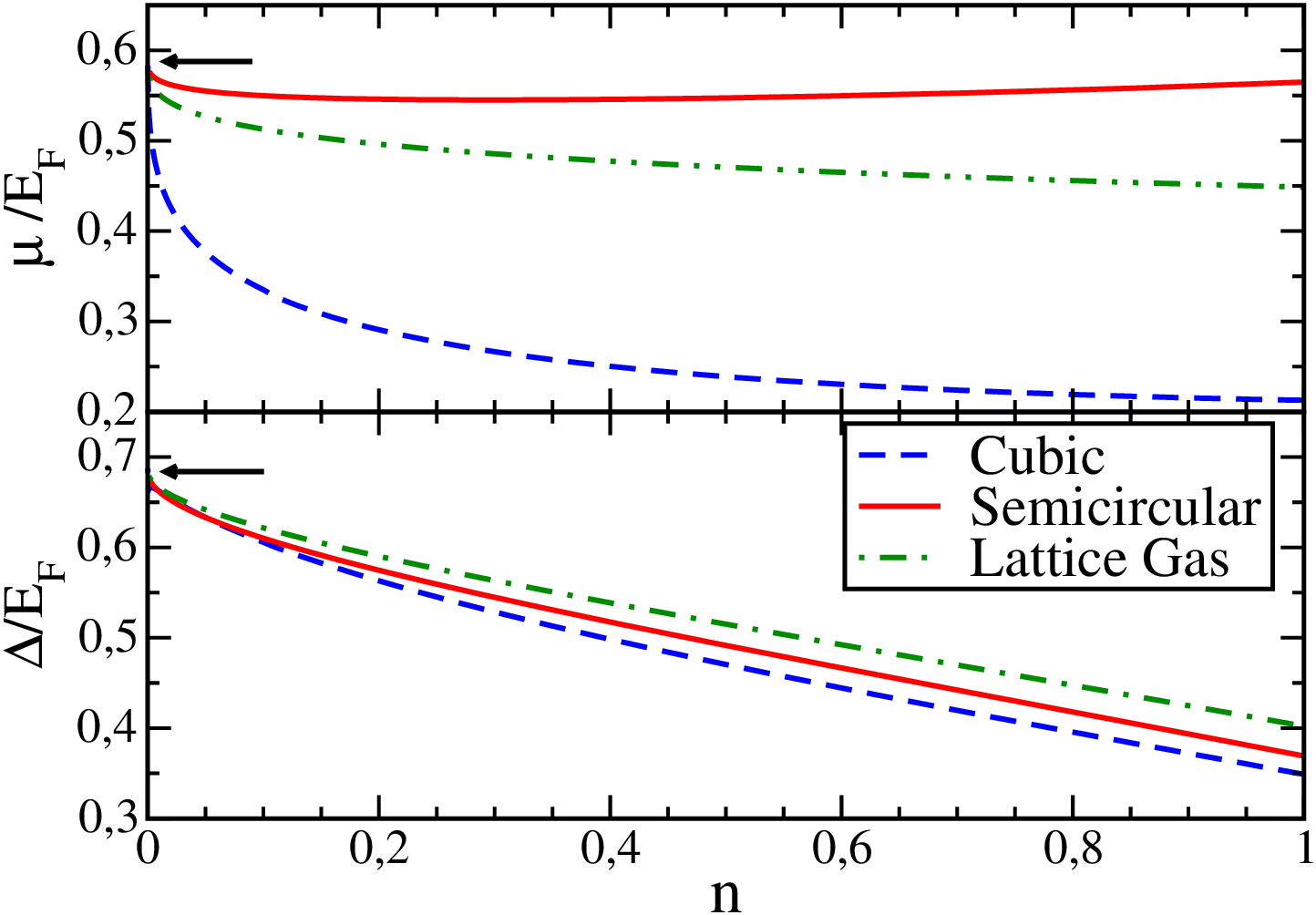}
\end{center}
\caption{\footnotesize{(Color Online) Static mean-field results for the 
rescaled chemical potential $\mu/E_F$ and superconducting order
parameter  $\Delta/E_F$ as a function of density $n$. The dashed blue
line is for the cubic lattice, the red solid line for the semicircular
DOS and the dot-dashed green line is for the lattice gas. 
Horizontal arrows mark Leggett's results in the zero-density limit.}}
\label{figure3}
\end{figure}

The equations correctly reduce to Leggett's results in the zero-density limit, where both the gap and the chemical potential are proportional to $E_F$.
 At finite density, corrections to the limiting value are introduced both by the different density of states with respect to the free DOS used by Leggett, and by the Hartree shift of the chemical potential.
The numerical solution of the MF equations (\ref{mf1},\ref{mf2}) for the reduced gap and chemical potential  is reported in Fig. (\ref{figure3}) for the three different lattices under considerations.
In order to highlight the effect of the lattice we plot the results from the dilute limit $n=0$ to the so-called half-filling ($n=1$) situation in which we have one fermion per lattice site.
The most evident result is the slowness of the evolution  from $n=1$, in which the effect of the lattice is maximum and the three models are obviously different, to $n=0$ where the models have to become equivalent.
The results appear to be strongly influenced by the $n=1$ value down to very low density, and the smoothness of the curves makes it questionable to define a reasonable low-density region from the data.
As a consequence of the persistence of the lattice effects, the finite-density behavior remains strongly dependent on the choice of the lattice. 

This effect is apparently much stronger on the chemical potential, while the superfluid order parameter depends in a weaker way on the choice of the lattice. This is most likely due to the fact that the unitary
regime is physically closer to the Bose-Einstein regime, in which the interaction strength controls the order parameter and the details of the bandstructure are less important, than to the BCS regime, where the DOS
 at the Fermi level has a major effect on the order parameter. Nonetheless, the dependence on density is still very strong for all the considered lattices.

In the case of the chemical potential (and of the renormalized internal energy, that we will show later in comparison with DMFT results), it is evident that the cubic lattice introduces significantly larger
 finite-density corrections at any density, while the two other models can give rise to smaller deviations. As mentioned above, this is essentially due to the peculiar value of $\tilde\mu$ at $n=1$ for the cubic lattice,
 ultimately arising from the peculiar behavior of the DOS. 
Finally, all the curves approach the zero-density limit from below and with a very large slope. This is a first indication that any attempt to extract informations about the dilute Fermi gas from a
lattice perspective needs a very careful extrapolation to the zero-density limit (besides the thermodynamic limit), and that the choice of the non-interacting density of states can be crucial to minimize these effects.

Another observation is in order. The behavior as a function of density of the rescaled quantities derives from a rather involved interplay between the ``large-density" effects 
(in which the precise choice of the lattice is very important) and the tendency towards the universal unitary limit. As a consequence, it is not possible to define a well-defined
 density scale under which the finite-density system already displays the universal physics. 
We can have regimes of density in which some quantities appear rather flat, or they have minima. In these regions or points one has that the derivative of the renormalized quantities as a function of the density is zero.
For the sake of definiteness, we have that, for the semicircular DOS one has $\partial \tilde\mu/\partial n = 0$, at $n \simeq 0.1$. If no data were available for smaller densities one could be tempted to interpret the vanishing derivative as the beginning of a regime in which $\tilde\mu$ is independent on $n$, i.e., a universal regime in which the dependence on density is lost.
Obviously, our MF data show that this is not the case, and the vanishing derivative is only associated to a minimum after which the dilute limit is approached. We observe that this could not be equally evident in a
simulation on a finite lattice, where only a finite number of densities can be considered.
In other words, the simple observation of a zero derivative of the observables as a function of density does not guarantee that the system is in the dilute, and authentically universal, regime, but rather it can suggest an artificial, or {\it fake} universality.

\section{Dynamical Mean-Field Theory of the Attractive Hubbard Model} 
\subsection{The General Formalism and the Model}
We briefly introduce DMFT and its application to attractive models and superfluidity. Previous DMFT studies of this  model have been so far mostly devoted  to the high-density regime. In Refs.\onlinecite{keller,caponeprluneg} the normal phase has been studied by excluding superfluidity, and a pairing transition between a normal metal and a pseudogapped state of preformed pairs has been found. Various properties of the superfluid state have been studied with the same approach, highlighting the ability of DMFT to properly describe the BCS-BEC crossover without bias towards one of the two limits\cite{randeria,toschi_njp,toschi_prb,hewson}.
The same approach has also been used to study two-component fermionic mixtures with mass unbalance\cite{dao_prb} and density unbalance.\cite{dao_prl}

The low-density regime of the attractive Hubbard model has been recently studied within DMFT in order to access the unitary regime in the normal state\cite{barnea_normal}
 and in the superfluid phase\cite{barnea}. In both papers, the cubic lattice has been used, and some limitations at low-densities have been reported. 
Here we extend the analysis of Ref. \onlinecite{barnea} considering different lattices, and we improve the numerical accuracy at low density, which will allow us to
 draw important conclusions about the ability of DMFT to describe the low-density limit.

DMFT maps a quantum lattice model onto a local problem, which can be represented through an ``impurity model''\cite{revdmft}, i.e., a model
 in which a single interacting site is embedded in a non-interacting medium.
In our case (\ref{hubbard}) is mapped onto an impurity model with attractive coupling, and the non-interacting bath is superfluid. Namely 
\begin{eqnarray}  
\label{aim}  
{\cal{  H}}_{AM} &  = &  \sum_{l,\sigma} \,   
\left[ \epsilon_l \, c^{\dag}_{l\sigma} \, c_{l\sigma} + V_l\,  
( c^{\dag}_{l\sigma} d_{\sigma} + \mbox{h.c.}) \right. \nonumber \\ 
&  + & \left. \Delta_l \,  
(c_{l\downarrow}^{\dag} c_{l\uparrow}^{\dag} + \mbox{h.c.}) \right] 
+ {\cal{H}}_{loc} 
\end{eqnarray}  
where $ {\cal{H}}_{loc} = -Un_{0\uparrow} n_{0\downarrow} - \mu n_0$ is the on-site term, and 
the chemical potential $\mu$ controls the density. For all the different lattices, $\mu=0$ corresponds to zero density
in the non-interacting case. 
From the impurity model we compute the normal and anomalous Green's functions, 
$G(\tau)=-\langle T c_{\uparrow}(\tau)  c^{\dag}_{\uparrow}(0)\rangle$ and   
 $F(\tau)=- \langle T c_{\uparrow} (\tau) c_{\downarrow}(0)\rangle$, and the corresponding normal and anomalous (superfluid) components of the self-energy $\Sigma$ and $S$.

The correspondence between the effective local model (\ref{aim}) and the original lattice model is guaranteed by a  self-consistency condition analogous to the Curie-Weiss equation for the Ising model.
 The self-consistency condition can be obtained by requiring that the  impurity Green's function coincides with the local component of the lattice Green's function, namely
\begin{eqnarray}
\label{selfcons}
G(i\omega_n)&= G_{latt}(i\omega_n) &\equiv\int_{-\infty}^{\infty}\hspace{-0.5cm} d\epsilon\  D(\epsilon)\  \frac{z^*-\epsilon}{\vert z-\epsilon\vert^2+S^2}
\nonumber\\
F(i\omega_n)&= F_{latt}(i\omega_n) &\equiv\int_{-\infty}^{\infty}\hspace{-0.5cm} d\epsilon\  D(\epsilon)\ \frac{-S(i\omega_n)}{\vert z-\epsilon\vert^2+S^2},
\end{eqnarray}
where $z=i\omega_n +\mu-\Sigma(i\omega_n)$.

We underline that the original lattice enters only through the non-interacting DOS, and that for the semicircular DOS (\ref{scdos}) this equation is particularly simplified, since it does not require the numerical integration on the energy $\epsilon$\cite{revdmft}.

It is important to  notice that DMFT can be also seen as an approximation of the full Luttinger-Ward functional in which only the local Green's function is considered\cite{revdmft}. This means that DMFT is a variational method. Since the static MF, in which the self-energy is local and it has no dependence on frequency, is a subcase of DMFT in which the self-energy loses the dynamical character, the variational principle implies that DMFT is always an improvement over static MF, unless the two methods become identical\cite{revdmft,variational}.  
   
\subsection{Exact Diagonalization Solution}
Despite the simplifications introduced in DMFT, the effective local model (\ref{aim}) can not be solved exactly by analytical methods and numerical solutions are necessary. Yet, the required computational effort is enormously lighter than for the original lattice model, and, more importantly, the method is defined in the thermodynamic limit, so that no finite-size corrections need to be considered, as opposed to full numerical solution of lattice models.  Analogously, we are in the grandcanonical ensemble, where we can tune the chemical potential and the density can assume arbitrarily small values.

Here we adopt Exact Diagonalization (ED)\cite{krauth}, which requires to truncate the sum  in Eq. (\ref{aim}) to a finite number of levels $N_s$. In practical implementations $N_s$ is necessarily small, but it has been shown that $N_s$ smaller than 10 provides accurate results for thermodynamic properties.\cite{krauth}  
Such a discretization introduces a new step in the iterative procedure. Namely, after a new bath is obtained by means of the self-consistency equation, it has to be represented into a discrete form. This representation is usually obtained by fitting the result of the self-consistency to a discrete system.  The details of the fit have to be chosen with care, especially in ``delicate" regions such as the low-density regime we are interested in.

As customary, we perform the fit on the imaginary-frequency axis (even if we work at $T=0$), where the Green's functions are smooth. This requires the definition of a fictive inverse temperature $\tilde\beta$ which defines a Matsubara grid. A large $\tilde\beta$ is required to investigate the low-frequency behavior because of the small energy scales. Here we used values of $\tilde\beta$ up to 6000 for the smallest densities. All the results we present are converged as a function of $\tilde\beta$. Other aspects of the fit are discussed, e.g., in Ref. \onlinecite{capone_1d,giorgio}.

ED has been used in the low-density limit of the attractive Hubbard model in Ref. \onlinecite{barnea}, where full diagonalization of the Hamiltonian matrix has been used, limiting the study to $N_s=6$ (5 levels in the bath).
Here we use the Lanczos algorithm, which allows to compute the groundstate for a larger $N_s$. We analyzed systematically the behavior of the results for  $N_s=6,7,8,9$, and we found that the results for static observables are essentially converged as a function of the number of bath levels for $N_s=8$, which is the value that we will use throughout the rest of the paper.

A direct measure of the systematic error associated with our ED solution is
the comparison between right-hand side and left-hand side of Eqs. (\ref{selfcons}).
In Ref. \onlinecite{barnea}, using up to $N_s=6$, the deviation between "lattice" and "impurity" estimates of the density, obtained by integrating over the frequency respectively the right-hand and left-hand sides of the first of Eq. (\ref{selfcons})  has been found to strongly increase as the chemical potential approaches the bottom of the band and the dilute region is approached.
 Here we found that a careful evaluation of Matsubara sums, together with the different values of $\tilde\beta$ and of the maximum
frequency used in the fit, can make the deviations much smaller than what found in Ref. \onlinecite{barnea} already for $N_s=6$. Increasing $N_s$ to 8 and 9, we have been able to reduce the difference between the two estimates down to 1\% at $n \simeq 0.002-0.003$, virtually eliminating discretization errors. Therefore the only limitation of our data is the DMFT approximation and no further uncertainty is introduced by the numerics.

\subsection{Zero density limit and Universal behavior}
\label{dmft_limit}

Even if we have almost eliminated the discretization errors, 
a more profound limitation appears in the zero density limit when the universal behavior is addressed. The key requisite to approach the unitary limit is the divergence of the s-wave scattering amplitude $a_s$. From a diagrammatic point of view $a_s$ is obtained as a ladder sum for the irreducible vertex part in the s-wave channel at $q=0$ in the vacuum. This leads to the simple condition for which $a_s$ diverges at the smallest interaction value for which the two-body problem develops a bound state.
Within DMFT the only contribution to the ladder sum comes from the local
vertex function\cite{revdmft}. It is straightforward to verify that this local approximation is not sufficient to recover the divergent $a_s$, but only leads to a finite value which depends on the chosen lattice. 
However, at static MF level in the broken-symmetry superfluid 
solution a static on-site vertex function $U=U_c$ is sufficient to
reproduce  the required criticality of the two-body problem, despite 
the non-divergent scattering length.
As a consequence of the finite ``local" scattering length $a_s^{loc}$, if we reduce the density 
maintaining the coupling strength at the ``universal value" $U_c$ for
which the full solution of the model would lead to a divergent $a_s$, we
will eventually enter a regime in which the quantity $k_Fa_s^{loc}$ decreases,
since the vanishing Fermi momentum is not compensated by the infinite 
scattering length. 
This will inhibit to reach a real universal regime beyond MF.
The necessity of non-local contributions in the zero-density limit
 is even more transparent if the symmetry is not broken (normal state). 
Here it is indeed necessary to include a nonlocal vertex function
(divergent in the ultraviolet in the continuum limit) in order to obtain
a finite contribution in the $k_F=0$ limit.
The above analysis implies that, while the DMFT calculations will be
accurate at finite density, when the density becomes extremely small the
approach will finally acquire a MF character with vanishingly small (weak-coupling) corrections. The value of the density below which this effect will take place is clearly a function of the effective  ``local" $a_s$, which measures how far from the unitary limit we are in practice. This limit density can be estimated by simply requiring that $k_Fa_s^{loc} \ll 1$. For the semicircular DOS we obtain a local scattering length $a_s^{loc} \simeq 1.08892$. This  implies that DMFT will reduce to static MF only for $n \ll  1.3 \cdot 10^{-3}$, a remarkably small value, smaller than the lowest densities attainable in present-time QMC simulations. 
Therefore DMFT is expected to be accurate down to extremely low density,
making it a reliable tool for the investigation of the small-density regime.

\section{DMFT Results}

Before discussing the role of the different lattices in the finite-density corrections to the universal limit, in this section we analyze the behavior of the most relevant observables in our DMFT solutions. For the sake of definiteness, in this section we limit ourselves to the semicircular DOS, which is the most easily implemented.

\vspace{1cm}
\begin{figure}[t]
\begin{center}
\includegraphics[width=7cm]{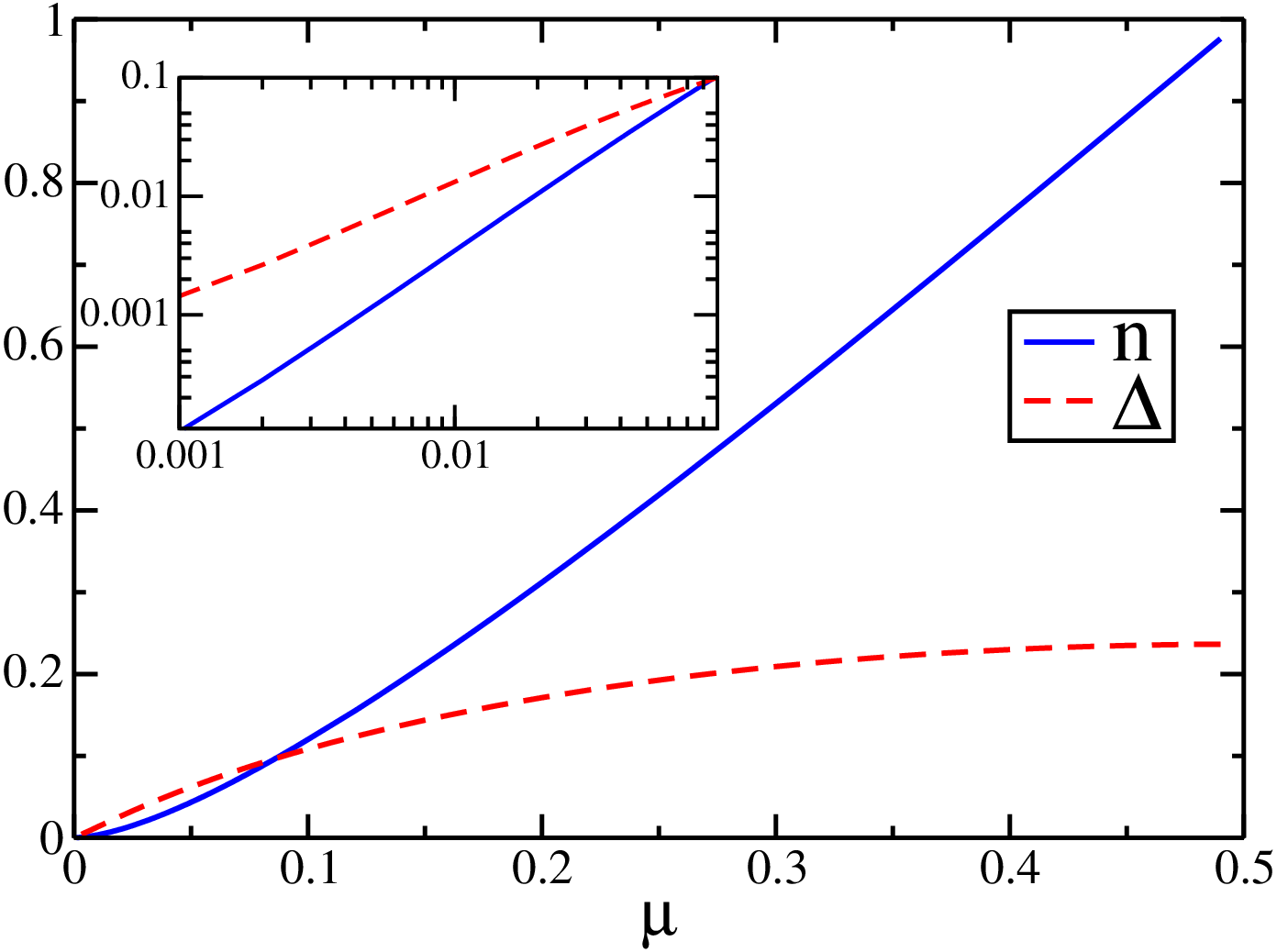}
\end{center}
\caption{\footnotesize{(Color Online) Particle density $n$ (solid blue line) and superfluid order parameter $\Delta$ (red dashed line) as a function of the chemical potential $\mu$ for a semicircular DOS.  The inset shows the low density region  in logarithmic scale, which shows that $n \propto \mu^{3/2}$ and $\Delta \propto \mu$.}}
\label{stateeq}
\end{figure}

As we already mentioned, DMFT is naturally expressed in the grandcanonical ensemble where the chemical potential is the natural variable. 
In Fig. \ref{stateeq} we plot $n$ and $\Delta$ as a function of $\mu$. Starting from half-filling ($n=1$), where the value of the chemical potential is fixed by the particle-hole symmetry of the model $\mu(n)_{|n=1}=D+U_c/2=D-D/2=D/2$, the density vanishes as the chemical potential vanishes, i.e. approaches the bottom of the band.
A first look at the $n(\mu)$ curve shows a rather smooth evolution from
a high-density region in which $n$ has essentially constant slope (i.e.,
the compressibility $\kappa = \partial n/\partial\mu$ is nearly
constant) to a low-density regime in which  $n \sim \mu^{3/2}$ or $\kappa \sim \mu^{1/2}$
 as expected from universality. The power-law
behavior is emphasized by the logarithmic scale used in the inset of Fig. \ref{stateeq}.
 Analogously $\Delta$ evolves from being nearly independent on $\mu$ for large densities to a linear behavior at small $\mu$. 

\begin{figure}[t]
\begin{center}
\includegraphics[width=7cm] {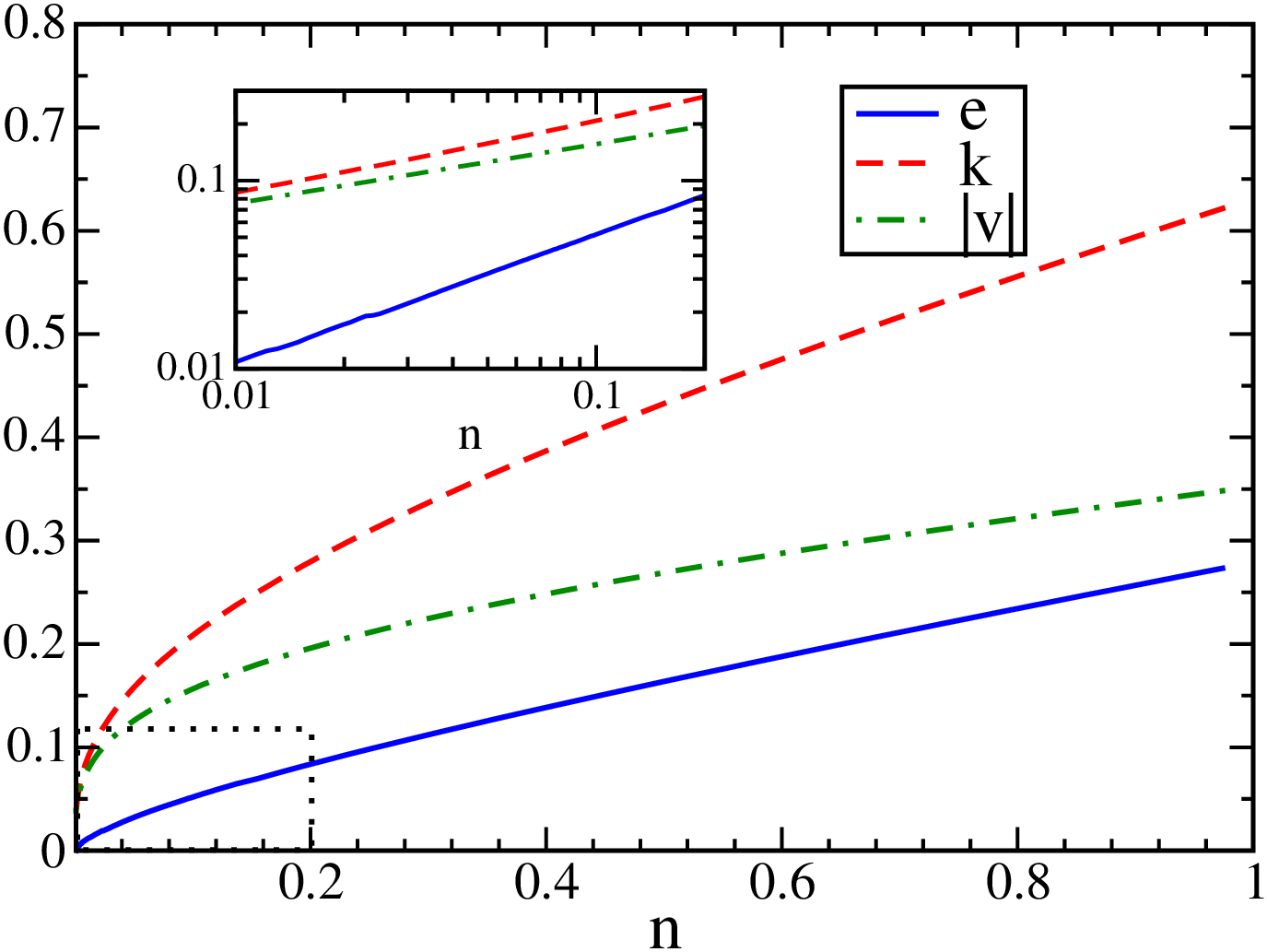}
\end{center}
\caption{\footnotesize{(Color Online) Ground state energy density (Total energy divided by the density) as a function of the density $n$.
We plot the total energy density $e$ (blue solid line), the kinetic contribution $k$ (red dashed line), the absolute value of the potential energy contribution $|v|$ (dot-dashed green line).
The inset shows the low-density region in logarithmic scale.}}
\label{figura4}
\end{figure}
Another relevant quantity in the unitary limit is the energy. To obtain the internal energy $E=V+K$, we computed the potential ($V$) and kinetic ($K$) energies. The potential energy per site is
\begin{equation}
  V= U \frac{1}{N}\sum_i\langle n_{i,\uparrow} n_{i,\downarrow}\rangle \equiv Ud,
 \end{equation}
where $N$ is the total number of lattice sites and $d$ is by definition the fraction of doubly occupied sites (local pairs) in the ground state, which can be computed directly as a static average of the double occupancy operator. The kinetic energy per site reads
\begin{equation}
\label{kinetic}
 K= \frac{2}{N}\sum_{\vec{k}\in FBZ} \epsilon_{\vec{k}} \langle n_{\vec{k}} \rangle =2T\sum_n \int_{-\infty}^{\infty} d\epsilon D(\epsilon)\epsilon G(\epsilon,i\omega_n)
\end{equation}
and it can be calculated from the lattice Green's function $G(\epsilon,i\omega_n)$. For the semicircular DOS 
Eq. (\ref{kinetic}) is simplified and it becomes a function of the local Green's functions only (even if a sum over Matsubara frequencies is still required):

\begin{equation}
\label{kinetic2}
 K = \frac{D^2T}{2}\sum_n [G(i \omega_n)^2 + G^*(i \omega_n)^2-2 F(i\omega_n)^2].
\end{equation}

We compared Eq. (\ref{kinetic}) and (\ref{kinetic2}) obtaining an almost perfect agreement, which allows us to confidently use Eq. (\ref{kinetic}) for the other dispersions in the following, for which the simpler relation (\ref{kinetic2}) does not hold. 

In Fig. \ref{figura4} we plot the energy per particle ¤$e=E/n$ together with the two contributions $k=K/n$ and $\vert v\vert = \vert V\vert/n$. The amplitude of the kinetic term is always larger than the potential term, correctly implying $E = K+V >0$, even if the difference rapidly shrinks as the density is decreased.
Interestingly, the two contributions both vanish as $n^{1/3}$ for small densities, as shown by the logarithmic plot in the inset of Fig. \ref{figura4}, while the total energy, obtained as their sum (or the difference between the two curves in Fig. \ref{figura4}), scales as $n^{2/3}$, the correct behavior for an energy-density in the dilute limit. 
These power-law behaviors have been recently proposed in Ref. \onlinecite{leggettnuovo} based on very general arguments, and are confirmed by our calculations. To our knowledge, this is the first numerical confirmation of these results. Moreover, the correct behavior of $e$, which emerges through a subtle cancellation of the leading contributions in $v$ and $\vert k\vert$, is a valuable confirmation of the accuracy of our estimates for the two contributions $K$ and $V$.

\subsection{Rescaled Quantities}

In this section we extend the MF analysis of Sec. \ref{meanfield_sec} computing the same observables using DMFT. As we will detail in the following, the dynamical nature of DMFT introduces sizeable improvements over static MF, but some general trends we discussed are found to be robust.

\vspace{1cm}
\begin{figure}[thb]
\begin{center}
\includegraphics[width=7cm] {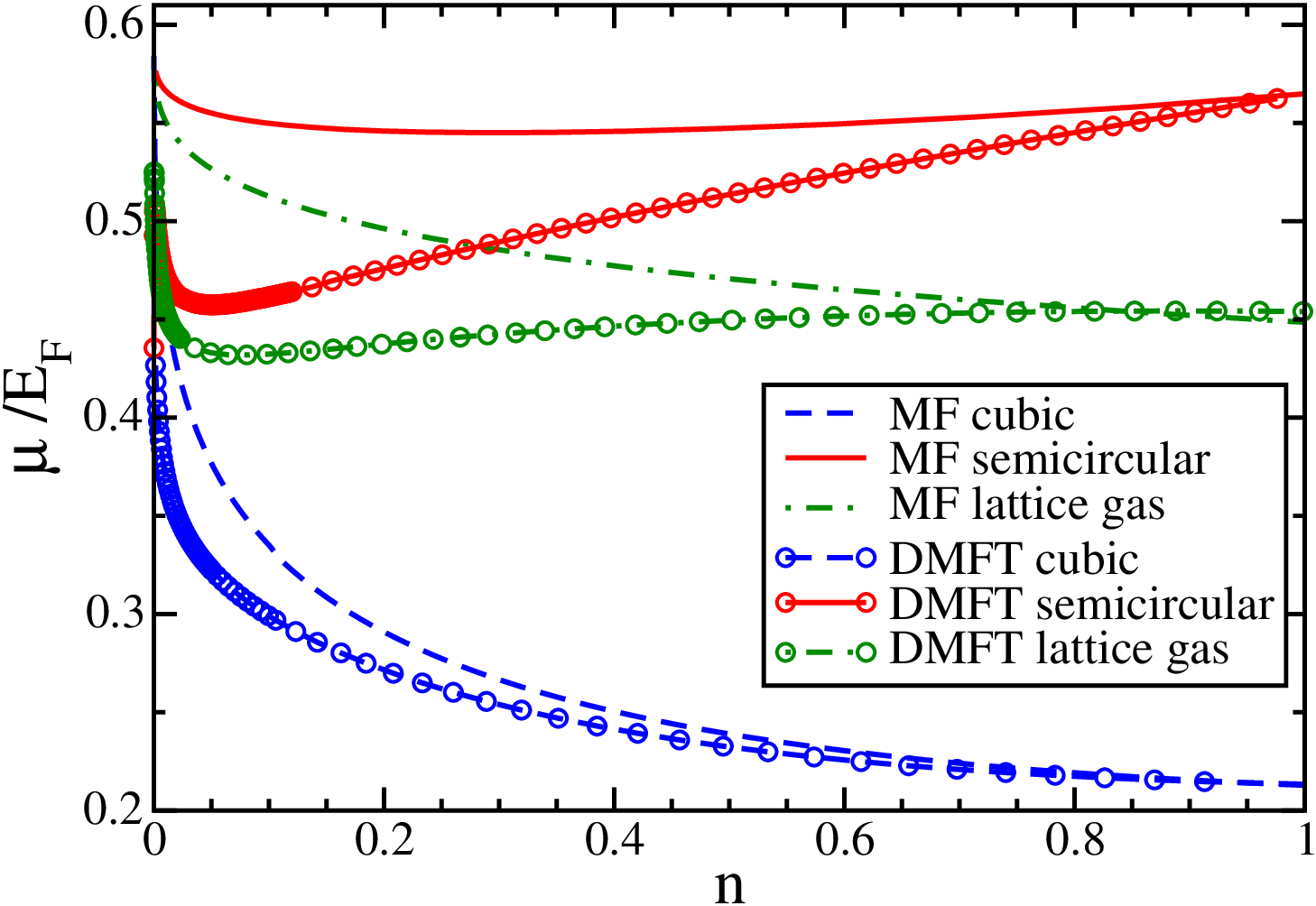}
\end{center}
\caption{\footnotesize{(Color Online) Renormalized chemical potential as a function of density. Comparison between static MF and DMFT (marked by open dots). Blue dashed line is for the cubic lattice, solid red line for the semicircular DOS and dot-dashed green line for the lattice gas.}}\label{compare}
\end{figure} 
\begin{figure}[thb]
\begin{center}
\includegraphics[width=7cm] {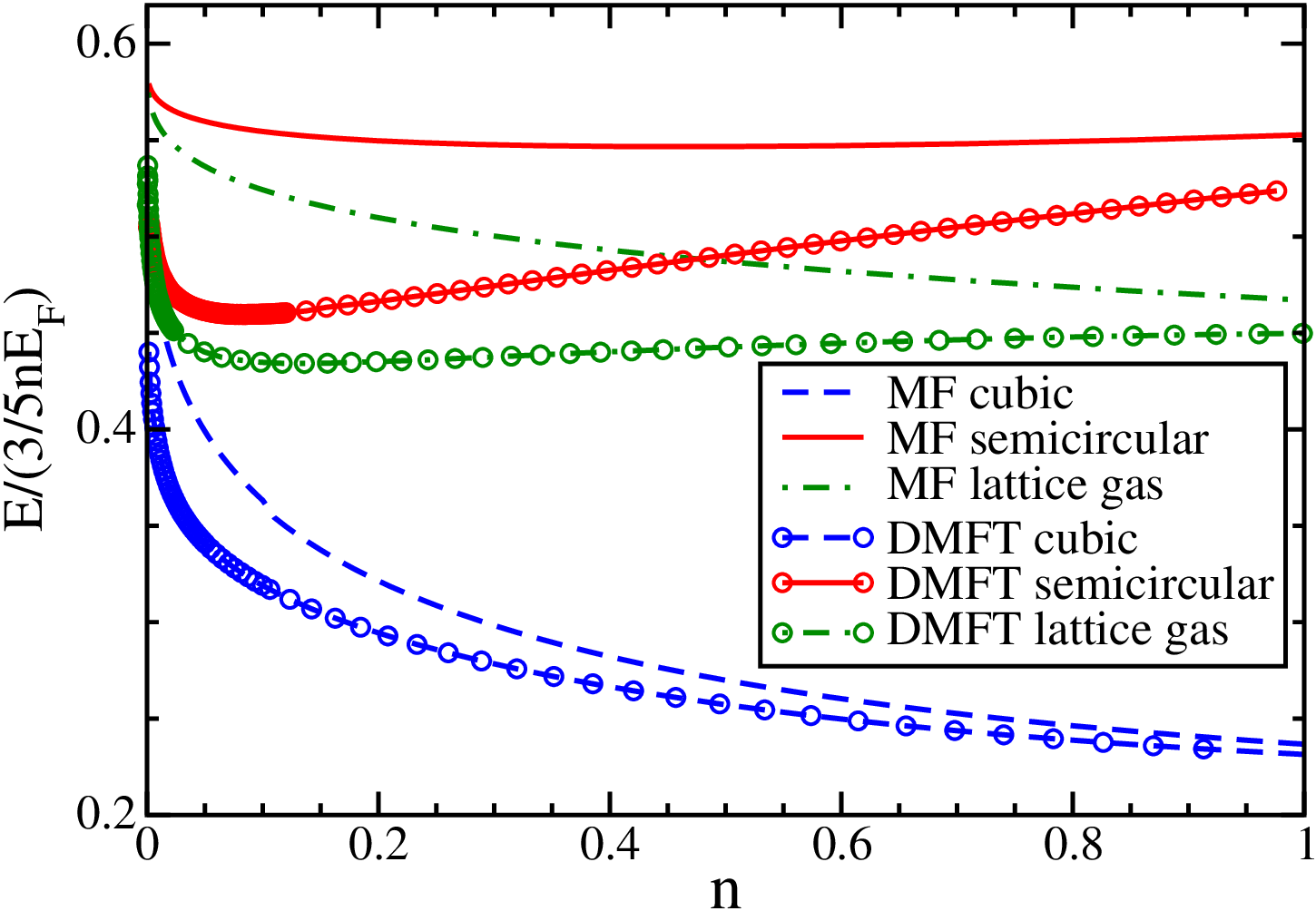}
\end{center}
\caption{\footnotesize{(Color Online) Renormalized internal energy as a function of density. Comparison between static MF and DMFT (marked by open dots). Blue dashed line is for the cubic lattice, solid red line for the semicircular DOS and dot-dashed green line for the lattice gas.}}
\label{compareenergy}
\end{figure}

\begin{figure}[t]
\begin{center}
\includegraphics[width=7cm] {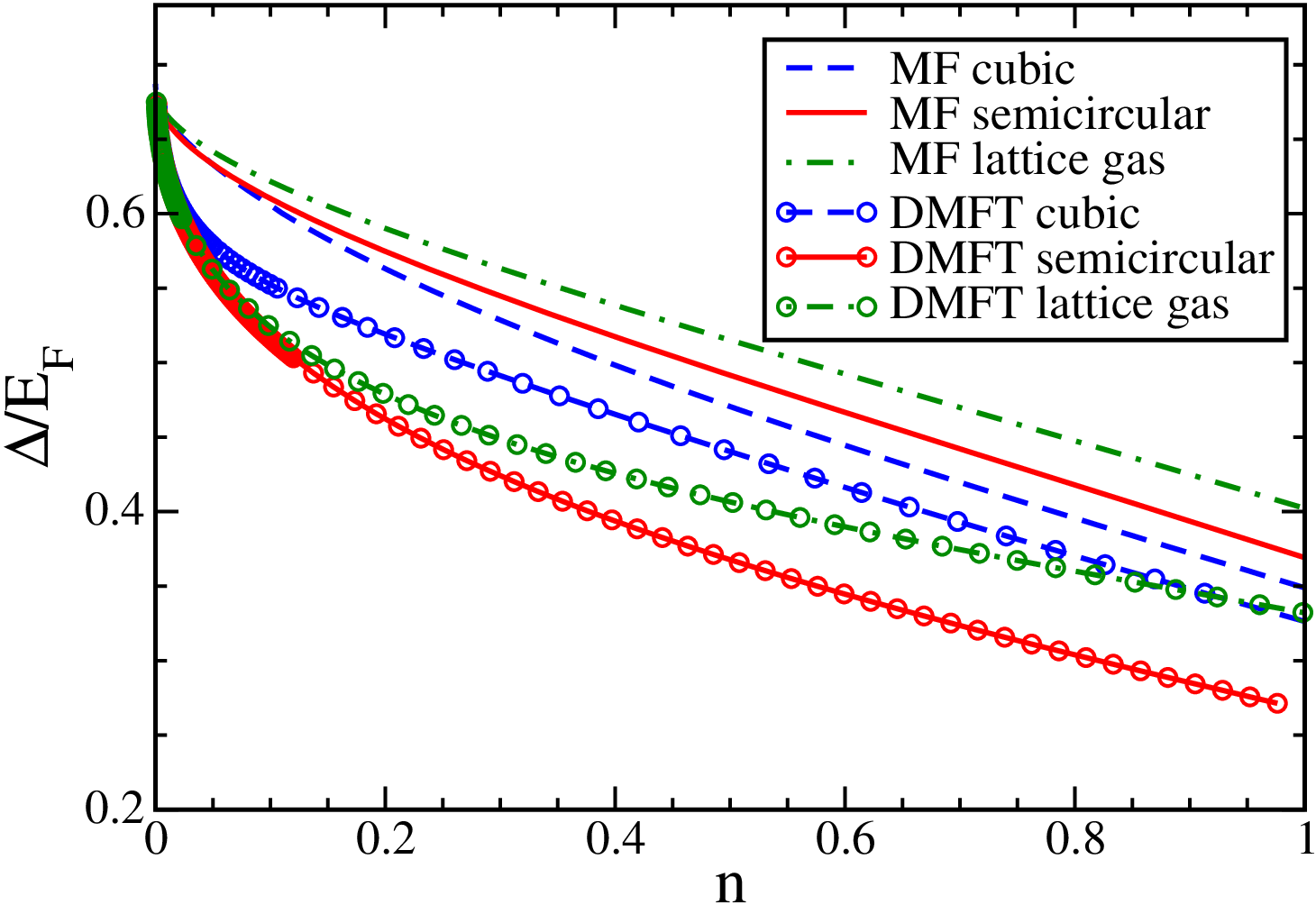}
\end{center}
\caption{\footnotesize{(Color Online) Renormalized superfluid order parameter as a function of density. Comparison between static MF and DMFT (marked by open dots). Blue dashed line is for the cubic lattice, solid red line for the semicircular DOS and dot-dashed green line for the lattice gas.}}
\label{comparedelta}
\end{figure}

The large correction introduced by DMFT with respect to MF is evident in Fig. \ref{compare},
\ref{compareenergy} and \ref{comparedelta}, where we compare,
respectively,  the reduced chemical potential $\mu/E_F$, superfluid
order parameter $\Delta/E_F$ and internal energy $\xi = E/(3/5nE_F)$
obtained in DMFT and in static MF for the three DOS's.
For $n=1$ the chemical potential is set by particle-hole symmetry, and
it is therefore the same in DMFT and MF\cite{notaparticlehole}. As soon
as the density is decreased, Fig. \ref{compare} clearly shows that DMFT
introduces a sizeable change with respect to MF, which is due to the
accurate treatment of quantum fluctuations. Interestingly, the
improvement brought by DMFT increases by reducing the density in a wide
range of densities, and it appears much larger for the semicircular and
lattice gas densities of states than for the cubic lattice. The lattice gas displays the weakest dependence on density.
 It is also evident, however, that the DMFT curves approach the same limit of the
static MF as the density actually approaches zero, 
as we expected from the arguments given in Sec. II, where we have shown 
that the including only local diagrams the scattering amplitude can not 
diverge and contributions beyond MF are bound to vanish for $n\to 0$.
The reduction of DMFT to static MF is confirmed by the behavior of the self-energies (not shown).
While for every finite density the self-energies have a non trivial frequency dependence, in the very
small density range in which DMFT rapidly collapses on the static MF, this frequency dependence
disappears and the self-energies become constants, as in static MF.
The results for the normalized internal energy $\xi = E/(3/5nE_F)$ are reported in Fig. \ref{compareenergy}, and show similar trends with respect to the chemical potential, confirming the anomalous behavior of the cubic lattice and the weaker density dependence of the lattice gas.

In Fig. \ref{comparedelta} we propose a similar comparison also for the
superfluid gap $\Delta$, again divided by the natural energy scale
$E_F$. As we have already discussed for the MF results, here the three
different lattices have a similar behavior. All the curves are
monotonically decreasing functions of the density, in contrast with 
the minima presented by the chemical potential and the energy in some of
the lattices. However, DMFT determines strong
corrections also for this quantity. It is intriguing that the different
lattices receive significantly different renormalizations. The cubic
lattice, that has the largest size effects in static MF is the one that
is less affected by the dynamical effects introduced by DMFT, while the 
semicircular and lattice gas DOS have similar shifts. In all cases the 
superfluid gap is reduced, as expected by the introduction of effects
beyond MF. Nonetheless, also for $\Delta$ the zero density limit becomes
identical to the static MF.

%\vspace{1cm}
%\begin{figure}[t]
%\begin{center}
%\includegraphics[width=7cm] {fig8.eps}
%\end{center}
%\caption{\footnotesize{(Color Online) Corrections introduced by DMFT on MF as a function of density for the three lattices.}}
%\label{differencesfig}
%\end{figure} 
%\vspace{1cm}

Nonetheless, it is indeed evident that, for all the observables,  the improvement brought by DMFT on MF increases as the density is reduced, until we reach a very low density for which the non-divergence of the scattering amplitude forces the result to rapidly approach the static MF. In this light, it is safe (and even prudentic) to assume that down to the densities at which the distance between MF and DMFT data is still increasing in Figs.\ref{compare},\ref{compareenergy},\ref{comparedelta}  DMFT data are not plagued by the lack of a divergent $a_s$. In the semicircular DOS and the lattice gas and for $\mu/E_F$ and $\xi$ this density scale is around $n \simeq 0.02$.

A clear outcome of our DMFT analysis is that the non-universal finite-density effects can be very large and they strongly depend on the choice of the lattice. For all lattices, the evolution from the large-density regime to the dilute regime is very smooth and regular.  The difference between the different lattices is still very large at $n \simeq 0.05\div 0.1$ or smaller, which are therefore not representative of  a real low-density regime where the physics becomes universal. This implies that any attempt to obtain informations about the universal regime should be based on a careful finite-density limit, and that densities much smaller than $n \simeq 0.05\div 0.1$ should be used in this extrapolation. Unfortunately modern QMC calculations are just at the limit of this
region (To our knowledge the smallest density used for the attractive Hubbard model is $n=0.05$ in Ref. \onlinecite{burovski3}), suggesting that further work is required to extract in a completely reliable way the properties of the universal Fermi gas.

A second important result is that the cubic lattice is the worst choice among the models we considered. In this model the large-density regime has a peculiar behavior which results from the singularities of the DOS, and it is very hard to wash out these effects by reducing the density. As a result, this model has the slowest convergence to the zero-density limit. Furthermore, the corrections introduced by DMFT are smaller than for the other lattices. 

The lattice gas and the semicircular DOS give rise to comparably accurate results, even if  the lattice gas presents a weaker dependence on the density, which makes it the best candidate for future studies. The correction introduced by DMFT on static MF is similar for the two models.

A direct comparison between our DMFT results and QMC simulations is not straightforward in light of the limits of DMFT in the zero-density limit. Nonetheless, if we simply consider the values of the observables around the density below which DMFT rapidly approaches the static MF, we find that the results of the lattice gas are in very good agreement with other estimates. For example for the lattice gas we have $\xi \simeq 0.43$ and $\tilde\Delta \simeq 0.55$ (the estimate for $\tilde\Delta$ is rather arbitrary, because this quantity does not show any minimum as a function of density). Comparing the latter two quantities with Fig. 14 of Ref. \onlinecite{bulgac2}, where several results from different numerical methods are collected, our data are clearly in the same range.

We conclude our analysis be presenting, as an example, the DMFT data for the chemical potential normalized by the {\it lattice} Fermi energy corresponding to each DOS. This shows that also the cubic lattice displays now a minimum as a function of density, which was not present in Fig. \ref{compare}. Indeed the comparison with the free-space Fermi energy is only meaningful at small density, especially for the cubic lattice, whose DOS is very different from the free one (see Fig. \ref{figura1}).
This plot shows that the existence of a minimum is a general results, which can be used to define a safe region in which DMFT is an improvement on static MF and an accurate description of a finite-density lattice model. This makes the argument about ``fake universality" that we discussed in Sec. \ref{meanfield_sec} relevant for any finite-density extrapolation to the zero-density limit.

\vspace{1cm}
\begin{figure}[t]
\begin{center}
\includegraphics[width=7cm]{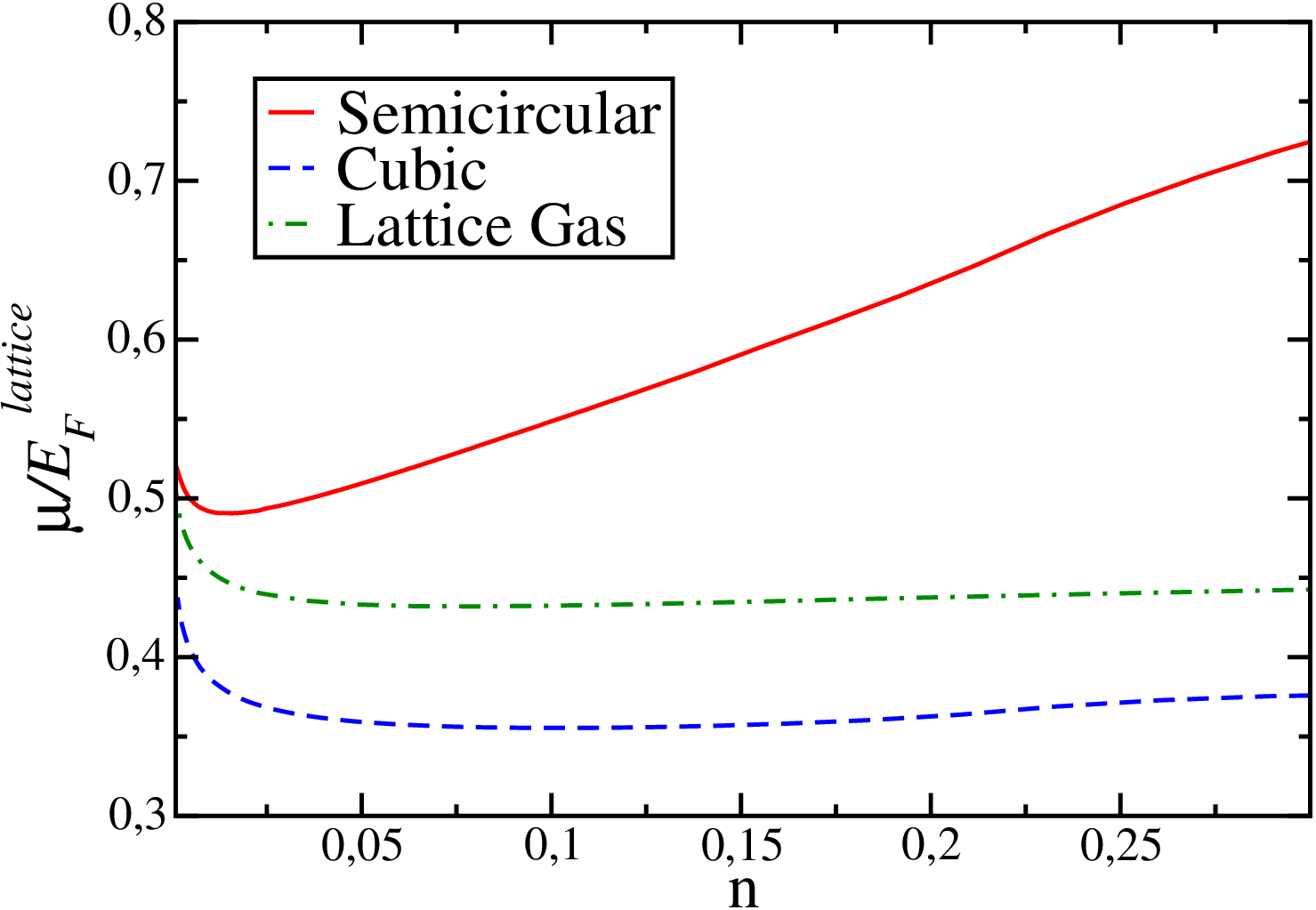}
\end{center}
\caption{\footnotesize{(Color Online) Dependence on the density of the chemical potential divided by the lattice free Fermi-energy for the three lattices under study}}
\label{differences2}
\end{figure} 
\section{Summary and Conclusions}

We have presented a thorough investigation of the approach to the zero-density limit of a lattice model with local attractive interaction. Three different lattice models have been considered, all sharing the low-energy density of states of free
fermions in three dimensions. Tuning the interaction strength at the unitary limit, for which the scattering length diverges, the three models approach the same zero-density limit, which is expected to reproduce the properties of a three-dimensional unitary Fermi gas. 

Our investigation is based on Dynamical Mean-Field Theory (DMFT), and for comparison on static mean-field at finite density. 
We show that DMFT introduces large corrections to MF down to very small densities, even if it is bound to reduce to static MF at zero density  because the local approximation inherent to DMFT does not allow for a divergence of the scattering length beyond MF. Following the evolution from large densities of the order of one fermion per site to zero density, the improvement introduced by DMFT increases as the density is reduced for a large window of densities. Only at very small densities of the order of $10^{-2}$ DMFT starts to approach static MF. 
The finite-density corrections are non-universal as expected and they turn out to strongly depend on the choice of the lattice. For all the lattice we considered, the evolution as a function of density is really smooth, and the large-density properties influence the physics down to very low densities. The three-dimensional cubic lattice, in particular, has the slowest convergence because of its peculiar DOS. The lattice gas and a system with a semicircular DOS display a much better convergence.

While the chemical potential and the energy strongly depend on the chosen lattice, the superfluid order parameter has a more lattice-independent behavior, even if the finite-density effects are sizeable. This can be associated to the fact that the unitary regime is closer to a strong-coupling Bose-Einstein regime, in which the interaction strength controls the order parameter and the details of the bandstructure are less important.

Our result highlight the importance of non-universal finite-density corrections to the unitary limit and impose serious constraint on any attempt to extrapolate the physics of the universal Fermi gas from finite-density calculations. In particular densities around $n \simeq 0.05 \div 0.01$ are not representative of the dilute regime, but they are still controlled by the large-density physics (and they strongly depend on the chosen lattice).

We finally remind that the limitations of DMFT are only dangerous when we want to describe the zero-density limit of a fermionic gas at unitarity. When we directly deal with lattice models, our results shown that the approach proves extremely accurate and introduces large corrections to static mean-field down to extremely low densities.


\begin{thebibliography}{99}
\bibitem{notaparticlehole} This is strictly true only for the semicircular DOS and for the cubic lattice. For the lattice gas the DOS is not symmetric and chemical potential may in principle be very different in different calculations. The very close results for DMFT and MF can be attributed to the existence of a sizeable gap, which strongly constraints the position of the chemical potential.
\end{thebibliography}
\end{document}